# Development of cloud, digital technologies and the introduction of chip technologies

Master of economics
Ali R. Baghirzade
Junior researcher of the research
Institute "Innovative Economics"
Plekhanov Russian University of
Economics
Moscow, Russia
e-mail:
fedorbagirov47@gmail.com

*Abstract* — *Hardly any other area of research has recently attracted as much attention as machine learning (ML) through the rapid advances in artificial intelligence (AI).*

*This publication provides a short introduction to practical concepts and methods of machine learning, problems and emerging research questions, as well as an overview of the participants, an overview of the application areas and the socio-economic framework conditions of the research.*

*In expert circles, ML is used as a key technology for modern artificial intelligence techniques, which is why AI and ML are often used interchangeably, especially in an economic context. Machine learning and, in particular, deep learning (DL) opens up entirely new possibilities in automatic language processing, image analysis, medical diagnostics, process management and customer management. One of the important aspects in this article is chipization. Due to the rapid development of digitalization, the number of applications will continue to grow as digital technologies advance.*

*In the future, machines will more and more provide results that are important for decision making. To this end, it is important to ensure the safety, reliability and sufficient traceability of automated decision-making processes from the technological side. At the same time, it is necessary to ensure that ML applications are compatible with legal issues such as responsibility and liability for algorithmic decisions, as well as technically feasible. Its formulation and regulatory implementation is an important and complex issue that requires an interdisciplinary approach. Last but not least, public acceptance is critical to the continued diffusion of machine learning processes in applications. This requires widespread public discussion and the involvement of various social groups.*

I. INTRODUCTION

In the modern world, digitalization is accelerating the pace of development. If about a century ago we lived in a time of industrialization, where the production of an innovative product took a long time, then now is the time for digital development, the innovations of which are manifested every day.

Due to the fact that this system has recently penetrated into human life and its nature has not been fully studied, and the digitalization process cannot be stopped, our main task is to properly consider the risks [1].

This article discusses both the pros and cons of digitalization and the statistics of the structure of IT services in Russian markets in all industries and at the household level.

## II. Statistic Data

Every year, the volume of the IT market in Russia and in the West, including Asian regions such as Japan, Singapore, China, Hong Kong, etc. grow.

Considering the example of Russia, the indicators in 2019 ($ 10,847 million) increased by 9.4% in relation to 2018 ($ 9,917 million), also in Figure 1, we can see a retrospective for 5 years, indicators for 5 years have grown by 41%.

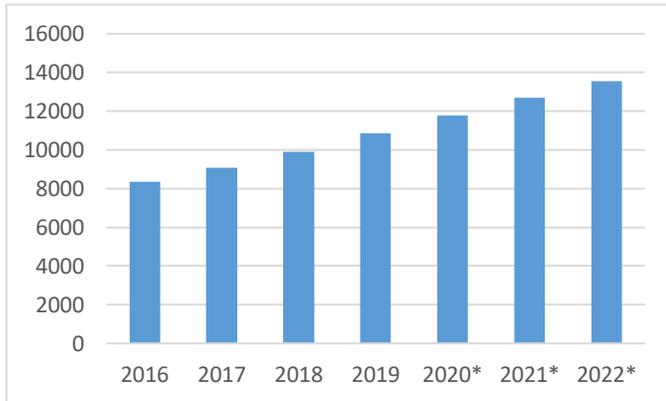

Figure 1. Development of the IT market in Russia

According to the forecast made by the authors, you can see that by 2022 in relation to 2019. The IT market will probably grow by 25%, which indicates a colossal rise in this market.

Manufacturing accounts for 25.1%, the public sector accounts for 21.4%, the banking sector occupies 20.9%, transport 4.3%, retail 5.2%, service and consumer 2.4%, telecom 12.1%, communicative services 6.8%, insurance 1.8% (Figure 2).

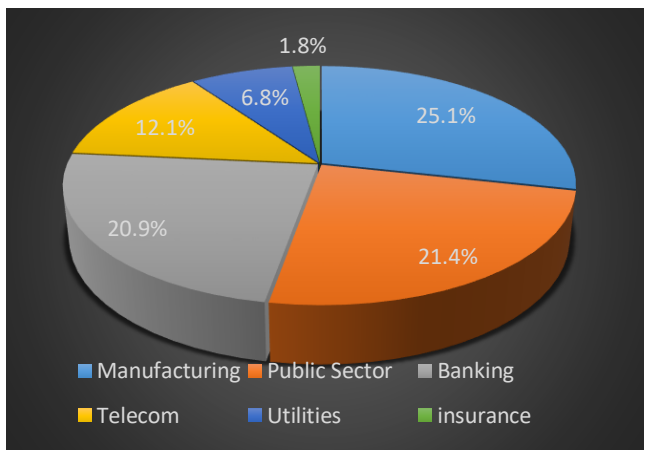

Figure 2. Market structure for 2019, RAS

Table 1

Market structure forecast for 2022, RAS

| Manufacturing | 26,7% |
|---|---|
| Public Sector | 20,8% |
| Banking | 21,1% |
| Telecom | 11,4% |
| Utilities | 6,7% |
| Insurance | 1,9% |

One of the most widespread and used devices at the moment is a smartphone. For 20 years, smartphones have evolved, becoming more accessible to most of the world's population.

According to a source from Bankmycell, smartphones are currently used by 45.12% (3.50 million people) of the world's population. 66.7% (5.17 million people) of the population use mobile phones

According to the forecast made by the author, in 2021 the growth of smartphone users will increase and amount to 3.8 million rubles. By 2025, the volume of users will be 57% (Figure 3).

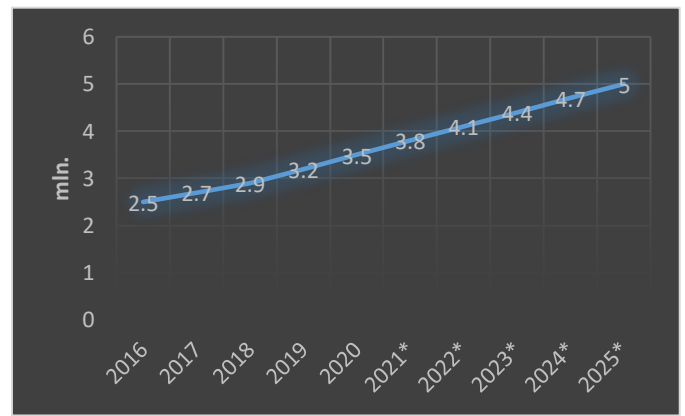

Figure 3. Population using smartphones (mln.).

As you can see in Table 2, the Chinese are the largest smartphone users, followed by the Indians and the third by the United States. The population in these regions is high, which speaks for itself. But there are also exceptions, for example Spain (№ 18, 46.4 million) and Nigeria (№ 23, 201 million).

Although Spain's population is lower than Nigeria, smartphone use in Spain is 15% higher. In this case, Spain is a developed state, and Nigeria is a developing state, respectively, per capita income in Spain is higher than in Nigeria, therefore, there are more smartphone users in Spain than in Nigeria (Table 2).

Table 2

Worldwide use of smartphones in the world.

| RANK | COUNTRY/MARKET | TOTAL POPULATION | SMARTPHONE USERS | SMARTPHONE PENETRATION |
|---|---|---|---|---|
| 1 | China | 1.42B | 851M | 59.9% |
| 2 | India | 137B | 346M | 25.3% |
| 3 | United States | 329M | 260M | 79.1% |
| 4 | Brazil | 212M | 96.9M | 45.6% |
| 5 | Russian Federation | 144M | 95.4M | 66.3% |
| 6 | Indonesia | 270M | 83.9M | 31.1% |
| 7 | Japan | 127M | 72.6M | 57.2% |
| 8 | Germany | 82.4M | 65.9M | 79.9% |
| 9 | Mexico | 132M | 65.6M | 49.5% |
| 10 | United Kingdom | 67.0M | 55.5M | 82.9% |
| 11 | France | 65.5M | 50.7M | 77.5% |
| 12 | Iran | 82.8M | 45.4M | 54.8% |
| 13 | Turkey | 83.0M | 44.8M | 54.0% |
| 14 | Vietnam | 97.4M | 43.7M | 44.9% |
| 15 | Philippines | 108M | 36.3M | 33.6% |
| 16 | South Korea | 51.3M | 36.1M | 70.4% |
| 17 | Italy | 59.2M | 36.0M | 60.8% |
| 18 | Spain | 46.4M | 34.5M | 74.3% |
| 19 | Pakistan | 205M | 32.5M | 15.9% |
| 20 | Bangladesh | 168M | 31.0M | 18.5% |
| 21 | Egypt | 101M | 30.5M | 30.1% |
| 22 | Thailand | 69.3M | 30.2M | 43.6% |
| 23 | Nigeria | 201M | 30.0M | 14.9% |
| 24 | Canada | 37.3M | 27.5M | 73.8% |
| 25 | Poland | 38.0M | 25.3M | 66.4% |
| 26 | Argentina | 44.6M | 23.6M | 53.0% |
| 27 | South Africa | 55.8M | 23.2M | 41.6% |
| 28 | Saudi Arabia | 33.3M | 22.7M | 68.3% |
| 29 | Malaysia | 31.5M | 20.9M | 66.5% |
| 30 | Colombia | 49.4M | 20.3M | 41.1% |
| 31 | Australia | 24.9M | 17.2M | 69.3% |
| 32 | Taiwan, China | 23.6M | 17.0M | 72.0% |
| 33 | Algeria | 41.7M | 15.8M | 37.9% |
| 34 | Morocco | 35.6M | 13.9M | 39.0% |
| 35 | Peru | 32.5M | 13.7M | 42.2% |
| 36 | Venezuela | 32.3M | 12.9M | 39.9% |
| 37 | Ukraine | 44.1M | 12.6M | 28.6% |
| 38 | Netherlands | 17.0M | 12.1M | 71.2% |
| 39 | Romania | 19.1M | 11.5M | 60.2% |
| 40 | Chile | 18.4M | 11.1M | 60.3% |
| 41 | Iraq | 39.7M | 9.6M | 24.2% |
| 42 | Belgium | 11.5M | 8.0M | 69.6% |
| 43 | Kazakhstan | 18.2M | 7.9M | 43.4% |

| RANK | COUNTRY/MARKET | TOTAL POPULATION | SMARTPHONE USERS | SMARTPHONE PENETRATION |
|---|---|---|---|---|
| 44 | United Arab Emirates | 9.5M | 7.8M | 82.1% |
| 45 | Sweden | 9.9M | 7.3M | 73.7% |
| 46 | Czech Republic | 10.5M | 7.1M | 67.6% |
| 47 | Azerbaijan | 10.0M | 6.9M | 69.0% |
| 48 | Portugal | 10.2M | 6.9M | 67.6% |
| 49 | Greece | 10.8M | 6.8M | 63.0% |
| 50 | Switzerland | 8.5M | 6.2M | 72.9% |

Considering the age category of the population, the main users of smartphones are people aged 18-29, the smallest, people aged 65+ (Table 3).

Table 3

Age category of using smartphones

|  | ANY CELL PHONE | SMARTPHONE ONLY | CELLPHONE ONLY (NON SMARTPHONE) |
|---|---|---|---|
| Total | 95% | 77% | 17% |
| Men | 95% | 80% | 16% |
| Women | 94% | 75% | 19% |
| Ages 18-29 | 100% | 94% | 6% |
| 30-49 | 98% | 89% | 9% |
| 50-64 | 94% | 73% | 21% |
| 65+ | 85% | 46% | 40% |

Exploring this topic, we can say that digitalization covers all areas of activity.

It is worth considering the field of medicine from a retrospective period.

### III. INTERNET TECHNOLOGIES AND THE IMPORTANCE OF CLOUD TECHNOLOGIES TODAY

In expert circles, machine learning is understood as a key technology in artificial intelligence. Artificial Intelligence is a branch of computer science that aims to enable machines to "intelligently" perform tasks. What does "smart" mean, and what technologies are used, is not specified. The first commercially significant advances in AI were made by so-called expert systems with a hand-built knowledge base. With this manual input of knowledge, or even with explicit programming of the solution, it is impossible to cope with more complex AI tasks. The alternative to this is machine learning, which is actually used today [2].

Represents the key technologies of intelligent systems. Machine learning seeks to derive "knowledge" from "experience" using learning algorithms to develop a complex model based on examples. The model, and therefore the automatically generated knowledge representation, can then be applied to new, potentially unknown data of the same type. When processes are too complex to analytically describe, but there is enough sample data available, such as sensor data, images, or text, machine learning is the ideal solution. The studied models can be used to make predictions or recommendations and decisions - without any rules or calculation rules set in advance.

Cloud computing is an IT infrastructure available, for example, over the Internet. It usually includes disk space, computing power, or application software as a service.

In more technical terms, cloud computing describes an approach to making IT infrastructures available over a computer network without having to be installed on a local computer.

These services are offered and used exclusively through technical interfaces and protocols such as a web browser. The spectrum of services offered in the context of cloud computing encompasses the entire spectrum of information technology and includes infrastructure, platforms and software, as well as storage space and databases [3].

The cloud is one of the oldest symbols of information technology and as such denotes computer networks, the interior of which is negligible or unknown.

Back in the early 1990s, some in the IT industry predicted that "computers will be distributed over the network," that is, that cloud computing will emerge as soon as the technology matures. The time came with the development of multi-tenant architecture in the late 1990s.

When the social network Facebook was launched in 2004, its members were given the opportunity to save and publish ("post") photos, videos, etc. on the Internet. However, the term "cloud computing" was mainly coined by some of the fast growing Internet companies such as Amazon, Google and Yahoo. Due to the rapid growth of their user base, these companies are faced with the problem of having to stock constantly growing systems that provide sufficient performance even during peak periods (for example, it would be a Christmas business for Amazon).

For Amazon, this peak load in 2006 was 10 times higher than the base load in daily work. To counter this problem, it was decided to transform the (service-oriented) architecture and services that were designed and built to cope with sometimes highly fluctuating or very large numbers of users into a product that was offered to the outside world. This means that this issue is spreading to cloud users during peak periods [4].

For Amazon, this move was a logical consequence in the early 2000s, as they had already moved internally to small, fast-paced teams (fast-paced "two-pizza teams") that had new features based on their existing cloud infrastructure.

implemented. Thus, the scaling effect of cloud services became the foundation of the cloud computing product itself, which has since been offered not only internally but externally. Amazon today is the world's largest cloud computing provider.

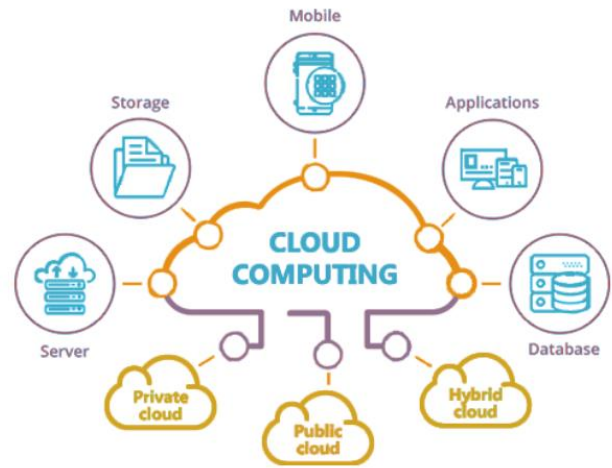

Figure 4. Cloud computing working

A prerequisite for using and distributing cloud computing services is broadband connections that are so fast that it no longer matters whether the data is stored locally on a PC or on remote servers in the cloud. Thus, the growing relevance of cloud computing for private users is associated with the offering of fast, reliable and cost-effective DSL and LTE connections to the market [5].

One of the most important technologies in the future that connects the network with the daily life of a person using cloud management technologies are chips. Chips are described in more detail in the next chapter.

IV. CHIPIZATION IN OUR DAYS

In 2001, ADS provided a 12 'Æ2.1mm VeriChip chip implant. This chip reads information up to six lines, plus a GPS is built into it. According to the developers, this function will help in case of missing people.

In 2017, in the city of Tyumen (Russia), a case was recorded when a student (Konstantin Polyakov), using an implanted chip, had the ability to pay for public transport.

Starting from July 17, 2003, the company (ADS Applied Digital Solutions) began its "chipization" in Mexico, more than

10,000 inhabitants of the country were "chipped" and 70% of clinics had devices that read information from these chips [6].

One of the amazing inventions of the 21st century is the 3D printer.

Israeli scientists from Tel Aviv University of Molecular Cell Biology and Biotechnology, using a 3D printer, printed a mini prototype of the heart, more suitable for a rabbit than for a person, but the human heart requires the same equipment and materials, consisting of adipose tissue, the hydrogel, which served as ink and cells.

As the head of the scientific group Dvir reported, "cells must create a pumping capacity in order to work together."

The researchers plan to first transplant 3D hearts into animals andif they succeed, the following step will be to carry out a similar operation on a human.

Dvir (head of the scientific group) also notes – "Perhaps within 10 years, when the organs will be printed, this methodology will be present in the best clinics in the world and the procedures will be carried out regularly".

The authors of the article - Detection and Control of Air Liquid Interface With an Open-Channel Microfluidic Chip for Circulating Tumor Cells Isolation From Human Whole Blood, proposed a bio-automation system for the isolation and recovery of circulating tumor cells (CTCs) individually from whole blood. To isolate CTCs, an approach based on open channel microfluidic chips is used. The proposed microfluidic chip design can form a stable air-liquid interface. MSW is trapped in the gaps between the supports of the microfluidic chip due to the capillary force associated with the meniscus at the air-liquid interface. The authors propose a microcircuit design to stabilize the air-liquid interface and sample flow rate. The authors present an image analysis algorithm for determining the position of the air-liquid interface. Using visual feedback from an image analysis algorithm, a control system is proposed to control the position of the air-liquid interface. The authors were able to stabilize the flow rate, which made it possible to complete the isolation of 5 ml of whole blood within 30 minutes. We have achieved an average error in the position of the air-liquid interface of 4 μm with a standard deviation of 7 μm. The authors confirmed that the position of the air-liquid interface is a decisive factor for the area of CTC capture. By controlling the position of the air-liquid interface, we achieved the capture of CTCs in a narrow band with high concentration [7].

In 2019, Elon Musk launched his new project called "Neuralink". This project involves the creation of implants, the electrode threads of which will be inserted and connected to the human brain using a neurosurgical robot. The first test of implantation of this chip with electrode filaments in the human brain is scheduled for the end of 2020.

The functions of this implant consist in that the people who undergo this operation are supposed to be able to use their brain to perform actions with electronic devices, make calls, and this will also give understanding in the treatment of diseases of the brain, spine, etc., but so far this activity has not been well studied. As Elon Musk suggests, this tool in the future may enter into symbiosis with artificial intelligence (AI) and the human brain, thereby leading to superhuman intelligence. Similar to this idea, scientists Roy Bucky and Philip Kennedy of Emory, Atlanta, implanted a 52-year-old John Ray with a microcircuit in his brain, thanks to which he could communicate and control electrical appliances. Such chips are already used to cure diseases such as: Parkinson's, epilepsy, sclerosis. These microcircuits will also have a memory department in which this or that information will be clearly stored, perhaps this is the future and the next stage of human evolution (Figure 5). According to the Author, each person has a unique DNA code, using these technologies, chips, it will be possible to accurately decipher the code, which in the future will allow humanity or, roughly speaking, cyberhumanity to heal and prevent 99.9% of all ailments of humanity, it is also possible a third branch will appear in the DNA code.

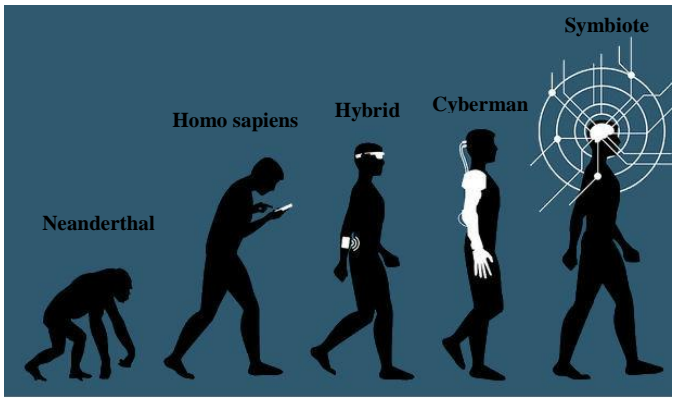

Figure 5. Human evolution, the era of the cyberman.

Considering the risks, we can say the following. It all depends on the human being. There is a possibility that data, built-in chips will suit someone without any complications, while someone may develop intolerance, allergies, rejection, thereby the body will turn on a defense mechanism, rejecting foreign material (body response). The risk also lies in the fact that there is a possibility that the implant will stop working, the chip will break. Then in these situations you cannot do without medical, surgical intervention.

In an interview with the head of the federal biomedical agency Veronika Skvortsova, the following is conveyed - Immune tests are a chip that can be used in clinics for multiple studies of the dynamics of a patient's condition, moreover, these chips allow one to predict and predict deterioration in health over a certain time.

## V. ABOUT TRANSPORT AND SERVICES

In our time, as mentioned earlier, digitalization is progressing. Soon we will not need such professions as cashier, taxi driver, pilot, machinist. Since, after eliminating all risks and repeated testing, drones (artificial intelligence) will take to the road, into the sky, underground.

In July 2018, the Yandex unmanned vehicle began to work like a real city transport - now residents of Innopolis in Tatarstan can use it for everyday travel. Today we signed an agreement on the creation of another test area for vehicle movement - on the territory of Skolkovo. There, the drone can be called directly in the Yandex.Taxi application.

Now the service of calling the drone is available to Skolkovo residents. Later it will become available to all guests of the innocentre. To use it, you need to accept the terms of the electronic agreement on participation in the testing of unmanned vehicles. You also need to be over 18 years old - our drone only drives adult passengers. The ride will be free, and the waiting time for the car depends on demand.

A self-driving car strictly observes traffic rules, detects and avoids obstacles, lets pedestrians pass and, if necessary, brakes urgently - in general, it can do everything that ordinary cars with a driver do. At the same time, we understand that drones are a new phenomenon on the roads. Therefore, in the cabin of our car there will always be a test engineer who monitors its movement (the authors of the unmanned vehicle transfer Yandex).

Self-driving cars are the future of urban transport. Transport companies will be the first to use unmanned technologies, including services for calling a taxi. And the opening of a test zone in the Skolkovo Technopark is another step towards the day when it will be possible to call an unmanned taxi in a couple of minutes in any city on the planet.

Uber has also launched a self-driving taxi project. But in 2018, during testing, a pedestrian was hit. The incident occurred due to the fact that a pedestrian crossed the road in the wrong place, so the artificial intelligence did not react in any way. As mentioned earlier, these risks need to be carefully considered in order to dispense with the above incidents in the future [8].

At this point in time, unmanned trains are already in use, for example, in countries such as Denmark, Spain.

Airplanes have also been on autopilot for a long time.

Many grocery stores already have terminals for paying for goods, in many cafes and restaurants, there are also terminals where you can make non-cash payments yourself.

Also in the banking and financial sector, digitalization is underway, people are slowly abandoning paper money and attach more importance to cards. Cryptocurrency is also widely used in our daily life [9].

## VI. Conclusion

When studying and analyzing this topic, the following conclusions can be drawn.

In general, experts highlight both the pros and cons of digitalization. While opening up new opportunities for communication, information exchange and acceleration of processes, digitalization, at the same time, in the near future can deprive a considerable number of people of their jobs, because many processes will be automated [10]. However, the opinion is expressed that in a detached perspective, as well as a result of the arrival of new technologies in the past, humanity will benefit more from these innovations, although there will be losers at a local moment. Reducing the need for human labor will further increase competition between people and will require improving professional skills.

**Acknowledgment**